\documentclass[journal=jacsat,manuscript=article]{achemso}
\setkeys{acs}{usetitle=true}

\usepackage[version=3]{mhchem} 
\usepackage{mciteplus}

\author{Katie E. Chong}
\affiliation{Nonlinear Physics Centre, Research School of Physics and Engineering, The Australian National University, 
	Canberra ACT 0200, Australia}
\author{Lei Wang}
\affiliation{Nonlinear Physics Centre, Research School of Physics and Engineering, The Australian National University, 
	Canberra ACT 0200, Australia}
\author{Isabelle Staude}
\affiliation{Nonlinear Physics Centre, Research School of Physics and Engineering, The Australian National University, 
	Canberra ACT 0200, Australia}
\alsoaffiliation{Institute of Applied Physics, Abbe Center of Photonics, Friedrich-Schiller-University Jena, 07743 Jena, Germany}
\author{Anthony James}
\affiliation{Center for Integrated Nanotechnologies, Sandia National Laboratories, Albuquerque, New Mexico 87185, USA}
\author{\\Jason Dominguez}
\affiliation{Center for Integrated Nanotechnologies, Sandia National Laboratories, Albuquerque, New Mexico 87185, USA}
\author{Sheng Liu}
\affiliation{Center for Integrated Nanotechnologies, Sandia National Laboratories, Albuquerque, New Mexico 87185, USA}
\author{Ganapathi S. Subramania}
\affiliation{Sandia National Laboratories, Albuquerque, New Mexico 87185, USA}
\author{Manuel Decker}
\affiliation{Nonlinear Physics Centre, Research School of Physics and Engineering, The Australian National University, 
	Canberra ACT 0200, Australia}
\author{\\Dragomir N. Neshev}
\affiliation{Nonlinear Physics Centre, Research School of Physics and Engineering, The Australian National University, 
	Canberra ACT 0200, Australia}
	\email{dragomir.neshev@anu.edu.au}
\author{Igal Brener}
\affiliation{Center for Integrated Nanotechnologies, Sandia National Laboratories, Albuquerque, New Mexico 87185, USA}
\author{Yuri S. Kivshar}
\affiliation{Nonlinear Physics Centre, Research School of Physics and Engineering, The Australian National University, 
	Canberra ACT 0200, Australia}

\title {Efficient polarization insensitive complex wavefront control using Huygens' metasurfaces based on dielectric resonant meta-atoms}

\abbreviations{IR,NMR,UV}

\begin{document}

\begin{abstract}
 Subwavelength-thin metasurfaces have shown great promises for the control of optical wavefronts, thus opening new pathways for the development of efficient flat optics.  In particular, Huygens' metasurfaces based on all-dielectric resonant meta-atoms have already shown a huge potential for practical applications with their polarization insensitivity and high transmittance efficiency.  Here, we experimentally demonstrate a polarization insensitive holographic Huygens' metasurface based on dielectric resonant meta-atoms capable of complex wavefront control at telecom wavelengths.  Our metasurface produces a hologram image in the far-field with 82\% transmittance efficiency and 40\% imaging efficiency. Such efficient complex wavefront control shows that Huygens' metasurfaces based on resonant dielectric meta-atoms are a big step towards practical applications of metasurfaces in wavefront design related technologies, including computer-generated holograms, ultra-thin optics, security and data storage devices.
\end{abstract}

The ability to perform wavefront control using optical metasurfaces has gained significant attention in recent years as it provides a route to ultra-thin optics that can potentially replace all bulky optical components~\cite{Yu2014a, Minovich2015, Genevet2015}. A number of studies have demonstrated the wavefront control capabilities of optical metasurfaces by realizing various flat optical elements, including beam deflectors~\cite{Yu2011, Lin2014, Yu2015}, beam shapers~\cite{Yu2011, Aieta2012a, Yang2014, Lin2014, Chong2015}, flat lenses~\cite{Hasman2003, Aieta2012a, Arbabi2015}, and holograms~\cite{Ni2013, Huang2013}.  Metasurface holograms, in particular, represent the ultimate examples of complex wavefront control, as they can transform an incident plane wave into a desired arbitrary wavefront in the far-field. As such, a number of groups have developed optical holographic metasurfaces~\cite{Ni2013, Huang2013, Zhou2013, Lin2013, Chen2014, Montelongo2014, Zheng2014, Zheng2015, Arbabi2015,Lalanne1998} aiming at applications such as holographic displays, data storage and optical tweezers. While borrowing from the vast knowledge of the field of computer generated holograms, metasurface holograms enable, for the first time, ultra-high efficiency of the image reconstruction based on a single-step lithographic process. Indeed, efficiencies of holographic reproduction in metasurfaces of more than 80\% have been demonstrated recently~\cite{Zheng2015, Arbabi2015a}.

However, most metasurfaces to date are based on plasmonic elements that are intrinsically lossy in the optical spectral region due to the absorption in metals~\cite{Yu2014a, Minovich2015, Genevet2015}. In addition, the need for the ability to control the optical phase-front in the full range of $0-2\pi$ imposes more restrictions on the design parameters of metasurfaces, therefore leading to the introduction of several undesirable losses, including reflection, diffraction and polarization conversion losses. These issues have triggered a rapid move towards all-dielectric metasurfaces~\cite{Lin2014, Chong2015, Arbabi2015, Decker2015, Staude2013}. While dielectric resonators in the microwave regime are known since the 80s \cite{Kajfez1986} and earlier, dielectric metasurfaces in the optical regime have attracted great interest only recently. By exploring resonant effects in dielectric nanostructures~\cite{Staude2013,Lin2014,Yu2015,Yang2014a,Chong2015,Arbabi2015,Bomzon2002,Levy2005,Yang2014,Wu2014,Chong2014,Horasaninejad2015,Iyer2015,Desiatov2015,Shalaev2015,Decker2015} that exhibit significantly lower losses than their metallic counterparts, such all-dielectric metasurfaces can offer practical solutions for a diverse range of efficient wave-shaping applications \cite{Decker2015,Chong2015,Yang2014,Arbabi2015}.  

Out of this large body of work, the concept of Huygens'~\cite{Pfeiffer2013} dielectric metasurfaces~\cite{Decker2015} has proven to be an important advance for achieving full phase control and high transmittance efficiencies. Huygens' metasurfaces rely on the overlap of the electric and magnetic resonances of the high-index dielectric nanoparticles \cite{Evlyukhin2011} to provide full phase coverage and complete elimination of reflection losses.  These particles furthermore show polarization insensitive operation~\cite{Chong2015, Yu2015} while, at the same time, absorption losses are negligible due to the use of lossless high-index dielectric materials.

Following this progress, recent works have demonstrated that all-dielectric Huygens' metasurfaces consisting of silicon nanoparticle arrays~\cite{Decker2015} enable efficient wavefront control, where the spatial variation of the phase can be achieved by tuning the lattice periodicity~\cite{Chong2015} or nanoparticle dimensions~\cite{Staude2013, Decker2015, Yu2015, Shalaev2015}. Such convenient control of the spatially dependent transmittance phase of the arrays, as required for wavefront shaping, has enabled the demonstration of some basic functionalities, including Gaussian-to-vortex beam shaping~\cite{Chong2015, Shalaev2015} and beam deflecting~\cite{Yu2015, Shalaev2015}.  However, simple wavefront control with a homogeneous phase profile (in either linear or azimuthal directions) across a metasurface can provide only limited functionalities.  

One of the main capabilities of metasurfaces is the on-demand generation of arbitrary wavefronts. An on-demand metasurface is able to imprint arbitrarily complex wavefronts on incident light with an inhomogeneous phase profile and thereby realize holographic images of any designs. Despite the desirability of such a metasurface, the well-sought capability of complex wavefront control has not yet been demonstrated with both high efficiency and polarization independence.

Here, we experimentally realize a dielectric holographic metasurface that produces an arbitrarily designed hologram image to demonstrate complex wavefront control.  Our holographic metasurface has a 82\% transmittance efficiency and a 40\% imaging efficiency, where the latter value represents the total amount of light that ends up in the pre-defined hologram pattern (see Method section).  These high efficiency values make our metasurface the most efficient polarization insensitive holographic metasurfaces to-date.

Additionally, while existing computer-generated-holographic technology can also produce arbitrarily designed hologram images, most of these holograms rely on diffractive optics that require multi-step lithography processes, which by itself could lead to alignment and other fabrication errors\cite{Freese2010,Scheuer2015}. Furthermore, conventional holography suffers from other disadvantages, including the twin image generation that fundamentally limits the hologram efficiency as well as the strong polarization sensitivity due to the holographic gratings. Here, our holographic metasurface presents a completely new approach to holography by using lossless resonant meta-atoms as Huygens' sources, fabricated in a single step in the lithography process. This opens the door to incorporating all types of metasurface functionalities as well as resolving problems with existing computer-generated-holograms.

\begin{figure}[th]
	\centering
	\includegraphics{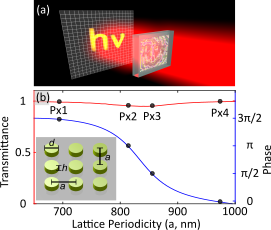}
	\caption{(a) A conceptual image of our study where a $\lambda=1477$~nm laser beam passes through a holographic metasurface to produce a two-dimensional hologram image behind the metasurface. (b) The simulated transmittance (red curve) and phase (blue curve) of silicon nanodisk arrays (schematic as inset) for a systematic variation of the lattice periodicity at a constant nanodisk diameter $d=534$~nm and height $h=243$~nm at $\lambda=1477$~nm.  Black dots denote the lattice periodicity values chosen for the four pixels used in experiment which cover a $3\pi/2$ phase range with close-to-unity transmittance.}
	\label{fig1}
\end{figure}

Figure~\ref{fig1}(a) shows a conceptual image of our experiment; a monochromatic laser beam ($\lambda=1477$~nm) passes through a holographic Huygens' metasurface which acts as a phase mask to modulate the wavefront of the incident beam, thereby generating a holographic image of the letters ``h$\nu$'' 12~mm behind the sample.

In order to experimentally realize the holographic metasurface, we employ silicon nanodisks as meta-atoms, which have previously allowed the realization of polarization insensitive beam-shaping metasurfaces with high transmittance efficiencies \cite{ Decker2015, Chong2015}.  All nanodisks forming the metasurface have the same diameter and the same height, and are embedded into a $580$~nm thick layer of silicon dioxide (silica). The height of the silicon nanodisks is set to $h=243$~nm. The diameter of $d=534$~nm has been chosen to tune the electric and magnetic dipole resonances into spectral overlap, which enables a $2\pi$ phase shift at the resonance \cite{Decker2015}, thereby providing a full range of accessible phase values. 

The spatial gradient of transmittance phases in the metasurface plane required for holographic image generation is obtained by the local variation of the lattice periodicity $a$ only. Since a variation of lattice periodicity would lead to a spectral shift of the resonance, the spatial transmittance phases introduced at a particular frequency would then be dependent on the range of lattice periodicity selected.  Our operation wavelength of 1477~nm is chosen such that the silicon nanodisk arrays with the selected parameters provide the largest range of transmittance phases for a physically realizable variation of lattice periodicity, as well as the highest transmittance across the same range of lattice periodicity values. Fig.~\ref{fig1}(b) shows the numerically calculated transmittance intensities and phases of the four optimized silicon nanodisk arrays. A schematic of an example array is shown in the inset of Fig.~\ref{fig1}(b).

The range of transmittance phases accessible by the variation of the lattice periodicity for our metasurface is $3\pi/2$, as shown in Fig.~\ref{fig1}(b), and is only limited due to our choice of using only one tuning parameter (the lattice periodicity $a$).  Although the maximum phase range can easily be extended by additionally changing the diameter and/or the height of the nanodisks, our design can already allow for a simple, discretized and pixelated implementations of the metasurface with equidistant phase steps, namely a four-level phase mask with phase steps of $\pi/2$.  The selected phase values are denoted by black dots on the blue curve in Fig.~\ref{fig1}(b),  showing the corresponding lattice periodicities of $a_{Px1}=695$~nm, $a_{Px2}=815$~nm, $a_{Px3}=855$~nm, and $a_{Px4}=975$~nm.  The close-to-unity transmittance of the nanodisk arrays with the selected periodicities is also shown as black dots on red curve in Fig.~\ref{fig1}(b).

\begin{figure}[th]
	\centering
	\includegraphics{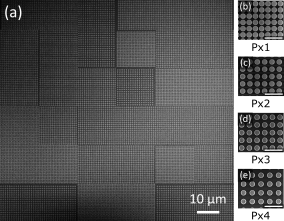}
    \caption{(a) Scanning-electron micrographs of a 6x6-pixel area of a typical hologram sample. (b)-(e) show magnified top views from typical individual pixels for the four realized lattice periodicities.  The scale bars denote the length of 2~$\mu$m.}
	\label{fig2}
\end{figure}

For fabrication of the silicon nanodisk metasurface, we use electron-beam lithography (EBL) on a back side polished silicon-on-insulator (SOI) wafer, followed by reactive-ion etching of the top silicon layer. The resulting nanodisks are then embedded in a 580~nm thick silica layer by low-pressure chemical vapor deposition (LPCVD). More details of the fabrication process can be found in the Method section.  Fig.~\ref{fig2}(a) shows the scanning-electron microscopy (SEM) image of a small section of a typical fabricated sample before LPCVD.  Figs.~\ref{fig2}(b)-(e) show magnified views of the structure from each of the four realized pixel types characterized by their lattice periodicities.

\begin{figure}[th]
	\centering
	\includegraphics{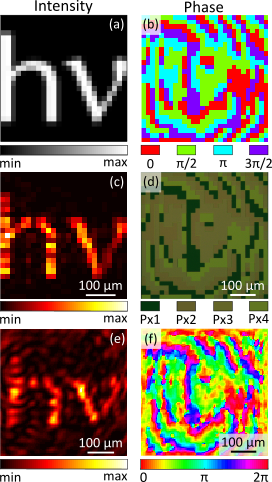}
	\caption{(a) Source image our hologram image is based on. (b)~Calculated phase pattern based on the source image. (c)~Simulated hologram image showing the expected hologram. (d)~Optical microscopy image of the fabricated holographic metasurface. (e) Experimental hologram image at 12~mm behind the sample plane with 40\% imaging efficiency.  (f) Phase reconstruction of the sample in the sample plane. }
	\label{fig3}
\end{figure}

For the arrangement of the pixels in the metasurfaces, we calculate the phase mask required for a particular input image using the Gerchberg-Saxton algorithm \cite{Gerchberg1972} and the angular spectrum method \cite{Matsushima2009}.  The calculation uses the source image shown in Fig.~\ref{fig3}(a), and several selected parameters, namely the resolution of the hologram image (28x28~pixels), pixel size (17.35x17.35~$\mu$m) and the distance of the hologram image from the phase mask (12~mm), to compute the phase mask pattern required to reproduce the source image as a hologram. These parameters are chosen in order to provide a high-quality hologram while at the same time keeping the EBL exposure time reasonably short. Naturally, larger writing fields (more pixels) allow for higher-resolution and more complex holograms. On the other hand, increasing the number of pixels under a restricted EBL writefield size of 500x500$~\mu$m can also enhance the hologram resolution.  However, minimizing the pixel size will affect the performance of the metasurface due to interparticle coupling effects or disturbances at the pixel borders, which are not taken into account in our design. Hence, the current pixel size of 17.35x17.35~$\mu$m was carefully chosen to balance all of these factors. 

Figure~\ref{fig3}(b) shows the calculated phase mask where the four distinct colors represent the phase shifts imprinted onto the incident wave in different spatial positions to generate the hologram image.  A simulated image (Fig.~\ref{fig3}(c)) of the reproduced hologram is also calculated by numerically propagating a 1477~nm plane wave through the phase mask pattern generated previously.  A true-color optical microscopy image at visible wavelengths of the fabricated phase mask, where the required phase shifts have been translated into silicon nanodisk arrays of the respective lattice periodicities, is shown in Fig.~\ref{fig3}(d), and can be compared directly with the original design in Fig.~\ref{fig3}(b).

To optically characterize our holographic metasurface phase mask, we first observe the generated hologram by transmitting a linearly polarized 1477~nm laser beam through the sample and imaging the hologram plane 12~mm behind the sample onto an Infrared (IR) camera.  The obtained image is shown in Fig.~\ref{fig3}(e), clearly showing the letters ``h$\nu"$, where the bright and dark features, as well as the size of the letters, agree very well with the calculated hologram in Fig.~\ref{fig3}(c).  Next, we measure the phase of the generated hologram by recording a set of four interferograms using a home-built Mach-Zehnder interferometer with the same laser source and IR camera as used for imaging. Based on these measurements, we perform a phase retrieval process \cite{Creath1988}.  The reconstructed experimental phase shown in Fig.~\ref{fig3}(f) is in good agreement with the calculated phase in Fig.~\ref{fig3}(b).  

While our holographic metasurface has been optimized for the operational wavelength of 1477~nm, this device can also operate and create visually similar hologram images within $\pm20$~nm of 1477~nm. This bandwidth of 40~nm is dependent on and limited by the resonance width of the dipole resonances.  Such dispersive behavior associated with the narrow resonances provides additional degrees of freedom for engineering optical responses. While a weakly dispersive metasurface will show a similar optical response for a broad range of frequencies, a resonant metasurfaces can in principle allow for tailoring a frequency selective response, e.g. displaying different holographic images for different colors.

\begin{figure}[th]
	\centering
	\includegraphics{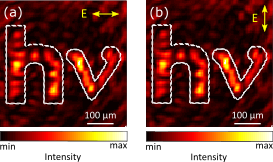}
	\caption{Hologram image generated with (a) horizontally- and (b) vertically-polarized light are visually identical, confirming the polarization-independence of our holographic metasurface.  The dash lines show the outline of the mask used to calculate the imaging efficiency.}
	\label{fig4}
\end{figure}

In order to provide a quantitative evaluation of the performance of the realized holographic metasurface, we measured the transmittance efficiency using an IR camera.  The transmittance efficiency of the holographic metasurface is measured to be 82\% for horizontal and vertical polarizations. Additionally, we measure the imaging efficiencies of the metasurfaces, \textit{i.e.} the total amount of light that ends up in the letters ``h$\nu"$, by masking the recorded hologram image as indicated by the dashed lines in Fig.~\ref{fig4}.  The imaging efficiency is measured as 39\% and 40\% for horizontal and vertical polarizations respectively.  Details of the measurement method and definitions of the efficiencies are explained in the Method section. 

Our imaging efficiency can be further improved by increasing the numbers of pixel of our hologram and reducing the pixel size. In this proof-of-principle experiment, the imaging efficiency and resolution of the hologram are limited by the writing time in the EBL process and our choice of source image resolution in order to produce a phase profile with more connecting pixels for each phase level to minimize stitching errors and proximity effects.  Further optimization of the hologram quality can be achieved for smaller pixel sizes, however large scale numerical simulations are needed to take into account the meta-atom interactions at pixel borders. Therefore, higher quality hologram images can in principle be achieved.

The visually identical hologram images recorded for the two polarizations, as shown in Figs.~\ref{fig4}(a) and (b) furthermore directly confirm the polarization insensitivity of the metasurface hologram, a property which cannot be achieved using holographic metasurfaces based on the acquisition of geometric phase.

In conclusion, we have demonstrated complex wavefront control using a highly efficient polarization-insensitive holographic Huygens' metasurface based on resonant silicon meta-atoms.  By varying only one geometrical parameter (lattice periodicity) that can be controlled easily during the fabrication process, we can effectively generate arbitrary hologram images with four phase-discretization levels.  The transmission efficiency and the imaging efficiency of our device is measured to be as high as 82\% and 40\% respectively. The functionality of our holographic Huygens' metasurface is polarization insensitive and therefore allowing it to be used in any optical system without pre-conditioning the incident beam, unlike metasurfaces that exploit geometric phase. Notably, by simply choosing an asymmetric design of the meta-atoms, one can also realize polarization sensitive holographic Huygens' metasurfaces, if required.  By further increasing the number of phase levels of the hologram and reducing the size of the pixels, the hologram efficiency, resolution, and complexity can be further increased, leading to potential applications in holographic displays, data storage and security devices based on computer-generated holography.

\section{Method}
\subsection{Fabrication}
For fabrication of the holographic metasurface, we performed electron-beam lithography (EBL) on a backside polished silicon-on-insulator wafer (243~nm top silicon thickness, 3~$\mu$m buried oxide thickness). We first cleaned the top silicon surface by oxygen plasma (2 min, 200 W), then spin-coated HMDS as an adhesion promoter (3000 rpm, 30 s) and followed by spin-coating of the negative-tone electron-beam resist NEB-31A(3000 rpm, 30 s).  The resulting resist thickness is about 300 nm. We performed both a pre-exposure bake (100~$^\circ$C, 2 min) and a post-exposure bake (90~$^\circ$C, 1 min). After electron-beam exposure (400 pA beam), development was performed using the MF-321 developer with a development time of 75 s, followed by a rinse in de-ionized water for several minutes. The resulting resist pattern was then used as an etch mask for an inductively coupled plasma (ICP) etching process (25~W RF power, 300~W ICP power, 60~$^\circ$C) using Ar (40 sccm) and HBr (15 sccm) as etch gases. Ar/Cl2 plasma chemistry is used for native oxide breakthrough. Remaining resist was removed using oxygen plasma and piranha solution. Finally, the sample is coated with a 580 nm thick silica layer by low pressure vapor deposition (LPCVD).

\subsection{Efficiency Measurement}
In order to provide a quantitative evaluation of the performance of the realized holographic metasurface, we derive the transmittance efficiency of the metasurface as the total intensity of the light going through the metasurface referenced to the intensity passing through an etched but unstructured area of the sample of equal size. This way, we eliminate the reflections from the handle wafer, which are not relevant for the performance of the actual metasurface.  

In addition to the transmittance intensity, we furthermore calculate the imaging efficiency, \textit{i.e.}, the total amount of light forming the pre-defined hologram pattern, the letters ``h$\nu"$, as recorded on the IR camera, divided by the amount of light passing through the same referenced area as in the transmittance efficiency case.  To calculate the intensity of the light that forms the letters ``h$\nu"$, we mask the recorded hologram image as indicated by the dashed lines in Fig.~\ref{fig4}. 

As the intensity distributions of the hologram in the image plane has a dynamic range exceeding that of the camera used in our experiments, multiple images were taken at different intensity levels of the incident laser beam using neutral density filters. This way it is possible to reconstruct the full dynamic range of the images to retrieve accurate intensity values and hence more accurate efficiency values.

\begin{acknowledgement}

This work was performed, in part, at the Center for Integrated Nanotechnologies, an Office of Science User Facility operated for the U.S. Department of Energy (DOE) Office of Science. Sandia National Laboratories is a multi-program laboratory managed and operated by Sandia Corporation, a wholly owned subsidiary of Lockheed Martin Corporation, for the U.S. Department of Energy's National Nuclear Security Administration under contract DE-AC04-94AL85000. 

K.E.C. acknowledges the support from the Australian Nanotechnology Network Overseas Travel Fellowships 2014 and the Australian National University Vice Chancellor's HDR Travel Grants 2014.  K.E.C., I.S., M.D., D.N. and Y.S.K. also acknowledge their participation in the Erasmus Mundus NANOPHI project, contract number 2013 5659/002-001. The authors also acknowledge a support from the Australian Research Council.

\end{acknowledgement}

%
%

\bibliographystyle{achemso}
\bibliography{hologram_ACS}

%
%
%
%
%

\end{document}